\documentclass[aps,twocolumn,superscriptaddress]{revtex4}
\usepackage{graphicx}

\begin{document}

\title{Switchable coupling for superconducting qubits using double
resonance in the presence of crosstalk}

\

\author{S. Ashhab}
\affiliation{Frontier Research System, The Institute of Physical
and Chemical Research (RIKEN), Wako-shi, Saitama 351-0198, Japan}

\author{Franco Nori}
\affiliation{Frontier Research System, The Institute of Physical
and Chemical Research (RIKEN), Wako-shi, Saitama 351-0198, Japan}
\affiliation{Physics Department and Michigan Center for
Theoretical Physics, The University of Michigan, Ann Arbor,
Michigan 48109-1040, USA}

\date{\today}

\begin{abstract}

Several methods have been proposed recently to achieve switchable
coupling between superconducting qubits. We discuss some of the
main considerations regarding the feasibility of implementing one
of those proposals: the double-resonance method. We analyze mainly
issues related to the achievable effective coupling strength and
the effects of crosstalk on this coupling mechanism. We also find
a new, crosstalk-assisted coupling channel that can be an
attractive alternative when implementing the double-resonance
coupling proposal.

\end{abstract}

\maketitle


Superconducting systems are among the leading candidates for the
implementation of quantum information processing applications
\cite{You1}. In order to perform multi-qubit operations, one needs
a reliable method for switchable coupling between the qubits, i.e.
a coupling mechanism that can be easily turned on and off. Over
the past few years, there has been considerable interest in this
question, both theoretically
\cite{Makhlin,Rigetti,Liu,Bertet,Paraoanu,Ashhab} and
experimentally \cite{Pashkin,Majer}.

The most obvious approach for switchable coupling is probably the
one based on the application of dc pulses \cite{Makhlin}. However,
given the difficulty of achieving accurate dc pulses and
complications associated with taking the qubits away from their
low-decoherence degeneracy points, recent efforts have focused on
proposals using fixed dc bias conditions, with the coupling being
turned on and off using ac signals
\cite{Rigetti,Liu,Bertet,Paraoanu,Ashhab}. The first such proposal
\cite{Rigetti} uses resonant oscillating fields applied to the two
qubits and employs ideas from the double-resonance concept in NMR
\cite{Slichter}. The method was generalized in Ref.~\cite{Ashhab}
to relax the condition of resonant driving on the two qubits.
Relaxing the resonant-driving requirement implies relaxing the
requirement of strong driving fields, a requirement that is
undesirable in superconducting qubit systems.

There is currently intense experimental effort to demonstrate
switchable coupling between superconducting qubits. Although the
apparently simple implementation of the double-resonance method
makes it an attractive option to use, a number of relevant
questions remain unanswered. In this paper we analyze some
considerations that would be useful to an experimentalist
attempting to implement it. We discuss mainly (1) the relation
between the strength of the applied fields and the resulting
effective coupling strength and (2) the possibility of reducing or
eliminating crosstalk effects using slow turn-on and turn-off of
the applied fields. The first question is important for deciding
the optimal driving parameters to use while taking into
consideration the limited decoherence times and limitations on the
strong driving of the qubits. In connection with the second
question, we find an alternative coupling channel that is easier
to drive and results in smoother oscillations.



{\it Switchable coupling using double resonance} -- We first
review, with a minor generalization, the derivation of the
double-resonance coupling mechanism of Ref.~\cite{Ashhab}. We
consider a system composed of two qubits with fixed bias and
interaction parameters. Oscillating external fields can then be
applied to the system in order to perform the different gate
operations.

The effective Hamiltonian of the system is given by:
\begin{equation}
\hat{H} = - \sum_{j=1}^{2} \left( \frac{\omega_j}{2}
\hat{\sigma}_z^{(j)} + \Omega_j \cos \left( \omega_j^{\rm rf} t +
\varphi_j \right) \hat{\sigma}_x^{(j)} \right) + \frac{\lambda}{2}
\hat{\sigma}_x^{(1)} \hat{\sigma}_x^{(2)},
\label{eq:Hamiltonian}
\end{equation}
where $\omega_j$ is the Larmor frequency of the qubit labelled
with the index $j$; $\Omega_j$, $\omega_j^{\rm rf}$ and
$\varphi_j$ are, respectively, the amplitude, frequency and phase
of the applied oscillating fields; $\lambda$ is the inter-qubit
coupling strength; and $\hat{\sigma}_{\alpha}^{(j)}$ are the Pauli
matrices with $\alpha=x,y,z$ and $j=1,2$ (note that we have set
$\hbar=1$). The eigenstates of $\hat{\sigma}_z$ are denoted by
$|g\rangle$ and $|e\rangle$, with
$\hat{\sigma}_z|g\rangle=|g\rangle$.

As discussed in Ref.~\cite{Ashhab}, we also take
$\lambda\ll\Delta\ll\omega$, where $\Delta=\omega_1-\omega_2$ (we
take $\omega_1>\omega_2$ and $\lambda>0$ with no loss of
generality), and $\omega$ represents the typical size of the
parameters $\omega_j$.

We now show how one can drive oscillations between the states
$|gg\rangle$ and $|ee\rangle$ using the double-resonance method.
We take the Hamiltonian in Eq.~(\ref{eq:Hamiltonian}) and
transform it using the unitary operation $\hat{S}_1(t) =
\exp\left\{i\sum_{j=1}^2 \left(\omega_1^{\rm rf}
\hat{\sigma}_z^{(1)} + \omega_2^{\rm rf} \hat{\sigma}_z^{(2)}
\right) t/2 \right\}$, such that we obtain the transformed
Hamiltonian $\hat{H}' = \hat{S}_1^{\dagger}(t) \hat{H}
\hat{S}_1(t) - i \hat{S}_1^{\dagger} (d \hat{S}_1/dt)$. Performing
a rotating-wave approximation (RWA), we obtain:
\begin{eqnarray}
\hat{H}' = & & \hspace{-0.3cm} - \sum_{j=1}^{2} \left(
\frac{\delta\omega_j}{2} \hat{\sigma}_z^{(j)} + \frac{\Omega_j}{2}
\left[ \cos\varphi_j \hat{\sigma}_x^{(j)} - \sin\varphi_j
\hat{\sigma}_y^{(j)} \right] \right) \nonumber \\
& & \hspace{-0.3cm} + \frac{\lambda}{4} \bigg(
\hat{\sigma}_x^{(1)} \hat{\sigma}_x^{(2)} \cos\delta\omega_{\rm
rf}t + \hat{\sigma}_y^{(1)} \hat{\sigma}_y^{(2)}
\cos\delta\omega_{\rm rf}t \nonumber \\ & & \hspace{0.5cm} +
\hat{\sigma}_y^{(1)} \hat{\sigma}_x^{(2)} \sin\delta\omega_{\rm
rf}t - \hat{\sigma}_x^{(1)} \hat{\sigma}_y^{(2)}
\sin\delta\omega_{\rm rf}t \bigg), \label{eq:Hprime}
\end{eqnarray}
\noindent where $\delta\omega_j=\omega_j-\omega_j^{\rm rf}$, and
$\delta\omega_{\rm rf}=\omega_1^{\rm rf}-\omega_2^{\rm rf}$. We
now perform a basis transformation in spin space from the
operators $\hat{\sigma}$ to the operators $\hat{\tau}$ such that
the time-independent terms in Eq.~(\ref{eq:Hprime}) are parallel
to the new $z$ axes and the new $y$ axes lie in the old $x-y$
planes (note here that we are performing two different
transformations for the two qubits). Following the same steps as
in Ref.~\cite{Ashhab} while choosing driving parameters that
satisfy the resonance condition
\begin{equation}
\sqrt{\delta\omega_1^2+\Omega_1^2} +
\sqrt{\delta\omega_2^2+\Omega_2^2} = \omega_1^{\rm rf} -
\omega_2^{\rm rf},
\label{eq:DoubleResonanceCriterion}
\end{equation}
\noindent we reach the effective Hamiltonian
\begin{equation}
\hat{H}'' = - \frac{\lambda_{\rm eff}}{2} \left\{ e^{i \delta
\varphi} \; \hat{\tau}_+^{(1)} \hat{\tau}_+^{(2)} + e^{-i \delta
\varphi} \; \hat{\tau}_-^{(1)} \hat{\tau}_-^{(2)} \right\},
\label{eq:Hdoubleprime}
\end{equation}
where we have defined
\begin{equation}
\lambda_{\rm eff} = \frac{\lambda}{4}
(1-\cos\theta_1)(1+\cos\theta_2),
\end{equation}
\noindent $\delta\varphi=\varphi_1-\varphi_2$,
$\tau_\pm^{(j)}=(\tau_x^{(j)}\pm i\tau_y^{(j)})/2$, and
$\tan\theta_j= \Omega_j/\delta\omega_j$. Alternative resonance
conditions are given by $\pm \sqrt{\delta\omega_1^2+\Omega_1^2}
\mp \sqrt{\delta\omega_2^2+\Omega_2^2} = |\omega_1^{\rm rf} -
\omega_2^{\rm rf}|$, and they result in simple variations of the
effective Hamiltonian given in Eq.~(\ref{eq:Hdoubleprime})
\cite{DevoretAcknowledgement}.

Equations (\ref{eq:DoubleResonanceCriterion}) and
(\ref{eq:Hdoubleprime}) form the basis for the double-resonance
coupling mechanism. The Hamiltonian $\hat{H}''$ drives the
transition $|gg\rangle\leftrightarrow |ee\rangle$ but does not
affect the states $|ge\rangle$ and $|eg\rangle$ in the basis of
the operators $\hat{\tau}$. Therefore, a single two-qubit gate
that can be performed using the Hamiltonian $\hat{H}''$ and the
set of all single-qubit transformations form a universal set of
gates for quantum computing.


{\it Effective coupling strength} -- The first question we
consider here is the relation between the strength of the driving
fields and the effective coupling strength, i.e. the two-qubit
gate speed. Equation (\ref{eq:DoubleResonanceCriterion}) can be
satisfied with values of $\Omega_j$ ranging from 0 to much higher
than $\Delta$, although our second RWA breaks down for those large
values (the deviations from the above derivation should be fixable
by including higher-order effects, as long as the driving
amplitudes remain much smaller than the qubit Larmor frequencies).

\begin{figure}[h]
\includegraphics[width=6.0cm]{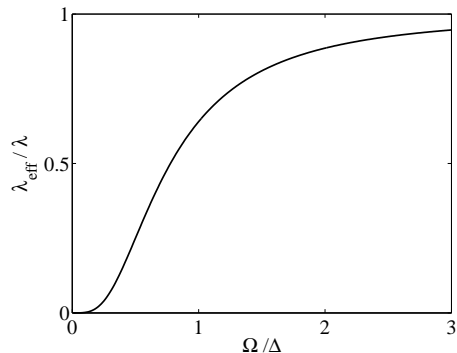}
\caption{The effective coupling strength $\lambda_{\rm eff}$
divided by the original coupling strength $\lambda$ from Eq. (1)
as a function of the driving amplitude $\Omega$ divided by the
inter-qubit detuning $\Delta$. Here we assume equal driving
amplitudes for the two qubits.}
\end{figure}

First let us look at the meaning of the coupling strength
$\lambda$. If the two qubits were exactly resonant with each other
($\omega_1=\omega_2$), two-qubit oscillations occur with (angular)
frequency $\lambda$. Although we are unaware of any mathematical
proof in the literature, it seems to us intuitively obvious that
this value should be the upper bound of how fast one can perform
two-qubit operations utilizing this coupling term. The ratio
$\lambda_{\rm eff}/\lambda$ therefore tells us how much of the
maximum speed we are achieving with a given set of driving
parameters. This ratio is given by:
\begin{equation}
\frac{\lambda_{\rm eff}}{\lambda} = \frac{1}{4} \left[ 1+
\frac{x-\frac{1}{4x}}{\sqrt{1+\left(x-\frac{1}{4x} \right)^2}}
\right]^2,
\end{equation}
where $x=\Omega/\Delta$, and we have assumed equal driving
amplitudes for both qubits. The ratio $\lambda_{\rm eff}/\lambda$
is plotted in Fig.~1. For weak driving fields, $\lambda_{\rm
eff}/\lambda$ is proportional to $\Omega^4$, making it highly
undesirable to go too far into that limit.

It might seem strange that it is possible to achieve the full
speed allowed by the coupling term in Eq.~(\ref{eq:Hamiltonian})
[see Fig.~1 with $\Omega/\Delta\rightarrow\infty$], a situation
intuitively reserved for the case of resonant qubits. Inspection
of the resonance condition (Eq.~\ref{eq:DoubleResonanceCriterion})
in this strong-driving limit shows that it corresponds to the
driving fields being far off resonance with the qubits and
\begin{equation}
\omega_1+\frac{\Omega_1^2}{2 \left( \omega_1 - \omega_1^{\rm rf}
\right)} = \omega_2+\frac{\Omega_2^2}{2 \left( \omega_2 -
\omega_2^{\rm rf} \right)}.
\end{equation}
The effect of the driving fields can therefore be understood in
terms of the ac-Stark shifts added to the qubit Larmor
frequencies. The resonance condition is satisfied when the
renormalized Larmor frequencies of the two qubits are equal. From
this point of view, it is not surprising that one can achieve the
full speed allowed by the coupling term \cite{geegOscillations}.
The experiment of Ref.~\cite{Majer} was close to this limit.

For any realistic situation, our approximations will break down
before achieving the infinitely strong driving limit. The
relationship between the different parameters determines how close
one can get to that limit. With the typical parameters
$\omega_1/2\pi = 5$ GHz, $\Delta/2\pi = 1$ GHz and $\lambda/2\pi =
0.1$ GHz, we numerically obtain 70-80\% of the maximum gate speed
before the dynamics deviates from that of
Eq.~(\ref{eq:Hdoubleprime}) [in all our numerical calculations, we
solve the Schr\"odinger equation using the Hamiltonian in
Eq.~(\ref{eq:Hamiltonian}). Here we also use the resonance
condition in Eq.~(\ref{eq:DoubleResonanceCriterion})].


{\it Crosstalk} -- We now address the question of crosstalk, i.e.
when each qubit feels the driving signal intended for the other
qubit. In order to describe the effects of crosstalk, we now add
to the Hamiltonian in Eq.~(\ref{eq:Hamiltonian}) terms of the form
$\beta \Omega_j \cos(\omega_j^{\rm rf}t+\varphi_j)
\sigma_x^{(j')}$, where $j\neq j'$, and the coefficient $\beta$
quantifies the amount of crosstalk. As the presently relevant case
for flux qubits, we take $\beta=1$ (i.e., 100\% crosstalk).

In order to understand the harmful effects of crosstalk
\cite{Vandersypen}, one should note that the Hamiltonian in
Eq.~(\ref{eq:Hdoubleprime}) does not operate in the computational
basis, but rather in a rotated basis. Reference \cite{Rigetti}
proposed using a precise timing procedure where meaningful results
are obtained only at times when the two bases coincide, whereas
Ref.~\cite{Ashhab} proposed applying the appropriate single-qubit
rotations before and after the two-qubit gate operation. Both
proposals add a step to the coupling procedure, and precise
calibration is required for this extra step. Corrections from
crosstalk must therefore be taken into account in implementing the
extra step.

Here we propose to turn the driving signal on and off slowly, such
that the adiabatic ramp of the driving signal's amplitude
transforms the states of the qubits between the computational
basis and the rotated basis of Eq.~(\ref{eq:Hdoubleprime}). With
this approach the extra steps mentioned above are no longer
needed. Clearly, and especially from an experimental point of
view, eliminating a step from the required procedure is always a
welcome simplification, regardless of crosstalk. With our new
proposal, one would need to adjust only one parameter in order to
obtain good oscillation dynamics.

Since the idea of slowly turning on and off near-resonant driving
fields has not been analyzed in the literature for this system, we
discuss it a little bit further here. We consider a single qubit.
We would like to take a qubit in an arbitrary quantum state and
adiabatically transform the quantum state from the computational
basis into the dressed-state basis, leaving the form of the
quantum state unaffected (note that the discussion of this system
is similar to that of a spin-1/2 particle in a changing magnetic
field). For a moment, we take the frequency of the driving field
to be fixed and the amplitude increasing from zero to its final
value. In this case we are effectively dealing with a
rotating-frame field that starts along the $z$ axis with initial
value $|\delta\omega|=|\omega-\omega^{\rm rf}|$ and changes (with
the additional component in the $x-y$ plane) to the final value
$\sqrt{\delta\omega^2+\Omega^2}$. In order to maintain
adiabaticity, the timescale over which we have to turn on the
driving fields must be larger than $1/|\delta\omega|$, since the
initial point is the most susceptible to non-adiabatic
transitions. In particular, note that in the case of resonant
driving (i.e., $\delta\omega=0$), it is impossible to turn on the
field adiabatically in the above sense. The main limitation
imposed by this adiabatic-turn-on approach is therefore that it
cannot be used with parameters that are too close to the
resonant-driving case ($\delta\omega=0$).

In the previous paragraph we have established the upper bound for
the allowed speed of turning on the driving fields, dictated by
single-qubit adiabaticity requirements. We now note that, in order
to maintain the switchable nature of the coupling mechanism, the
process of turning on the driving field must be fast compared to
the timescale of two-qubit oscillations. Using the same parameters
as above ($\omega_1/2\pi = 5$ GHz, $\Delta/2\pi = 1$ GHz and
$\lambda/2\pi = 0.1$ GHz), we find that $|\delta\omega|$ must be
substantially larger than 25 MHz in order to be able to satisfy
both requirements. If we take $|\delta\omega|$ to be a few hundred
MHz, we find that we have a large enough window of values for the
turn-on time such that this process is slow with respect to the
single-qubit dynamics, but fast with respect to the two-qubit
dynamics.

\begin{figure}[h]
\includegraphics[width=8.0cm]{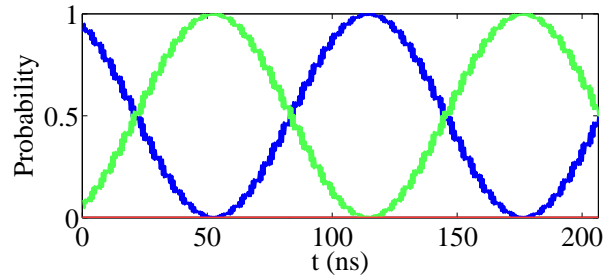}
\caption{(color online) The occupation probabilities of the four
computational states as functions of time. The black, blue (gray),
green (light-gray) and red (dark-gray) lines (the black and red
lines are indistinguishable from the $x$ axis) correspond,
respectively, to the states $|gg\rangle$, $|ge\rangle$,
$|eg\rangle$ and $|ee\rangle$. The initial state is $|ge\rangle$,
$\omega_1/2\pi$=5 GHz, $\omega_2/2\pi$=4 GHz, $\lambda/2\pi$=0.1
GHz, and $\delta\omega_1=-\delta\omega_2$=0.15 GHz. The amplitudes
of the driving fields were adjusted to reach optimal results,
keeping $\Omega_1=\Omega_2$. The driving fields are turned on and
off over durations of 20 ns (note that the time of turning on and
off the driving fields is not included in the time axis). In order
to see the reduction in crosstalk effects, this figure can be
compared with Fig.~4(b) of Ref.~\cite{Ashhab}.}
\end{figure}

Figure 2 shows oscillations that characterize a two-qubit gate
using driving fields that are turned on and off slowly as
explained above. Comparison with Fig.~4(b) of Ref.~\cite{Ashhab}
shows that the crosstalk-induced, noise-like fluctuations are
greatly reduced. Note that the oscillations are between the states
$|ge\rangle$ and $|eg\rangle$, in contrast to the appearance of
Eq.~(\ref{eq:Hdoubleprime}). The difference is due to the
different single-qubit rotations performed in the pulsed and
adiabatic approaches.


Interestingly, crosstalk opens a new possibility for coupling
between the qubits. An alternative resonance condition is
(approximately) given by
\begin{equation}
\sqrt{\delta\omega_1^2+\Omega_1^2} =
\sqrt{\delta\omega_2^2+\Omega_2^2}.
\label{eq:CrossTalkDoubleResonanceCriterion}
\end{equation}
This can be seen by inspecting Fig.~1 of Ref.~\cite{Ashhab}. This
resonance condition corresponds to oscillations between the states
$|gg\rangle$ and $|ee\rangle$, using the adiabatic-ramp approach
above \cite{EarlierMistake}. In the absence of crosstalk, one can
say that the effective matrix element for this coupling channel
vanishes, and no coupling occurs even if the resonance condition
is satisfied. In the presence of crosstalk, two-qubit oscillations
do occur.

\begin{figure}[h]
\includegraphics[width=8.0cm]{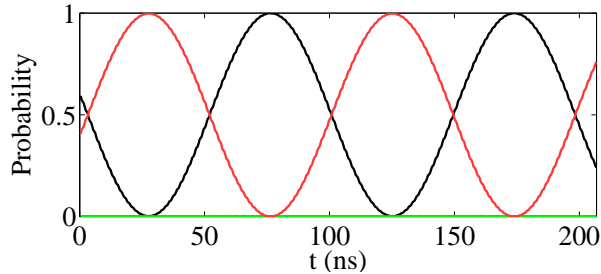}
\caption{(color online) Same as in Fig.~2 (the blue and green
lines are now indistinguishable from the $x$ axis), but the
initial state is $|gg\rangle$ and $\delta\omega_1=$0.14 GHz.}
\end{figure}

This coupling mechanism can also be treated analytically as above.
However, the algebra is now more cumbersome and the results less
transparent. The implementation procedure, however, is
straightforward. One could for example set the qubit-field
detunings $\delta\omega_1>0$ and $\delta\omega_2<0$, with a small
difference between $\delta\omega_1$ and $|\delta\omega_2|$. One
also sets $\Omega_1=\Omega_2$ [this choice means that
Eq.~(\ref{eq:CrossTalkDoubleResonanceCriterion}) cannot be
satisfied; however, crosstalk shifts modify the resonance
condition such that we obtain a resonance peak]. By varying this
common amplitude, one can easily locate the resonance. An example
of this coupling mechanism is shown in Fig.~3. An important
advantage of this approach is that for weak driving the gate speed
is higher and the resonance peak is broader in this case than in
the traditional case presented above, allowing for more error
tolerance using this coupling mechanism (of the order of 5-10\%
compared to 0.5\% for typical parameters). The resulting
oscillations are also smoother in this case.


{\it Conclusion} -- We have analyzed some considerations related
to the possible experimental implementation of the
double-resonance method for achieving switchable coupling between
superconducting qubits. We have obtained the dependence of the
effective coupling strength on the amplitudes of the applied
driving fields. We have also discussed the approach of using slow
turn-on and turn-off of the driving fields. This approach reduces,
and possibly eliminates, the effects of crosstalk. The two main
questions addressed in this paper are important in determining the
optimal set of parameters for experimental attempts to implement
this coupling mechanism. We have also found that crosstalk opens a
new coupling channel that has some advantages over the one used in
the original idea of the double-resonance approach.

We would like to thank K. Harrabi, J. R. Johansson, J. Martinis,
Y. Nakamura and G. S. Paraoanu for useful discussions. This work
was supported in part by the National Security Agency (NSA), the
Army Research Office (ARO), the Laboratory for Physical Sciences
(LPS), and the National Science Foundation (NSF) grant
No.~EIA-0130383. One of us (S.A.) was supported by the Japan
Society for the Promotion of Science (JSPS).


\begin{thebibliography}{99}
\bibitem{geegOscillations} Note here that in this case $\theta_2=\pi$, so
that the coupling oscillations occur between the states
$|eg\rangle$ and $|ge\rangle$, in spite of the appearance of
$\hat{H}_{\rm eff}$ in Eq.~(\ref{eq:Hdoubleprime}).
\bibitem{EarlierMistake} One must also be careful to avoid satisfying
this resonance condition if one wishes to use the coupling
mechanism mentioned earlier in the manuscript. These
crosstalk-assisted oscillations were mentioned, but erroneously
dismissed as being harmless, in Ref.~\cite{Ashhab}.
\end{thebibliography}
\end{document}